\documentclass[preprint]{aastex}

\usepackage{psfig}

\newcommand{\hanii}{H$\alpha+$[\ion{N}{2}]} 
\newcommand{\oiii}{[\ion{O}{3}]}
\newcommand{\oii}{[\ion{O}{2}]}
\newcommand{\ha}{H$\alpha$}
\newcommand{\hb}{H$\beta$}
\newcommand{\nii}{[\ion{N}{2}]}
\newcommand{\sii}{[\ion{S}{2}]}
\newcommand{\oi}{[\ion{O}{1}]}
\newcommand{\neiii}{[\ion{Ne}{3}]}
\newcommand{\nev}{[\ion{Ne}{5}]}
\newcommand{\feii}{[\ion{Fe}{2}]}

\begin{document}

\title{Near-Infrared Observations of Powerful High-Redshift Radio \\
Galaxies: 4C~40.36 and 4C~39.37 \altaffilmark{1,2} }

\author{E.\ Egami\altaffilmark{3,4}, L.\ Armus\altaffilmark{5}, 
G.\ Neugebauer\altaffilmark{3,4}, T.\ W.\ Murphy Jr.\altaffilmark{3,6},  
B.\ T.\ Soifer\altaffilmark{3,5}, K.\ Matthews\altaffilmark{3}, and
A.\ S.\ Evans\altaffilmark{7}
}

\altaffiltext{1}{Some of the data presented herein were obtained at
the W.\ M.\ Keck Observatory, which is operated as a scientific
partnership among the California Institute of Technology, the
University of California, and the National Aeronautics and Space
Administration.  The Observatory was made possible by the generous
financial support of the W.\ M.\ Keck Foundation.}
\altaffiltext{2}{Based in part on observations made with the NASA/ESA
Hubble Space Telescope, obtained at the Space Telescope Science
Institute, which is operated by the Association of Universities for
Research in Astronomy, Inc., under NASA contract NAS~5-26555. These
observations are associated with proposal \#7860.}
\altaffiltext{3}{Palomar Observatory, California Institute of Technology,
320-47, Pasadena, CA~91125}
\altaffiltext{4}{Current address: Steward Observatory, University of
Arizona, 933 North Cherry Avenue, Tucson, AZ 85721-0065}
\altaffiltext{5}{SIRTF Science Center, California Institute of
Technology, 314-6, Pasadena, CA~91125}
\altaffiltext{6}{Current address: Department of Physics, University of
Washington, Box 351560, Seattle, WA 98195}
\altaffiltext{7}{Department of Physics and Astronomy, State University
of New York at Stony Brook, Stony Brook, NY~11794-3800}
\begin{abstract}

We present near-infrared imaging and spectroscopic observations of two
FR~II high-redshift radio galaxies (HzRGs), 4C~40.36 ($z=2.3$) and
4C~39.37 ($z=3.2$), obtained with the Hubble, Keck, and Hale
Telescopes.  High resolution images were taken with filters both in
and out of strong emission lines, and together with the spectroscopic
data, the properties of the line and continuum emissions were
carefully analyzed.  Our analysis of 4C~40.36 and 4C~39.37 shows that
strong emission lines (e.g., \oiii\ 5007 \AA\ and \hanii) contribute
to the broad-band fluxes much more significantly than previously
estimated (80\% vs. 20--40\%), and that when the continuum sources are
imaged through line-free filters, they show an extremely compact
morphology with a high surface brightness.  If we use the
$R^{1/4}$-law parametrization, their effective radii ($r_{e}$) are
only 2--3 $h_{50}^{-1}$ kpc while their restframe $B$-band surface
brightnesses at $r_{e}$ are $I_{e}(B) \sim$ 18 mag/$\sq$\arcsec.
Compared with $z \sim 1$ 3CR radio galaxies, the former is $\times$
3--5 smaller, while the latter is 1--1.5 mag brighter than what is
predicted from the $I_{e}(B)-r_{e}$ correlation.  Although exponential
profiles produce equally good fits for 4C~40.36 and 4C~39.37, this
clearly indicates that with respect to the $z\sim1$ 3CR radio
galaxies, the light distribution of these two HzRGs is much more
centrally concentrated.  Spectroscopically, 4C~40.36 shows a flat
($f_{\nu}$=const) continuum while 4C~39.37 shows a spectrum as red as
that of a local giant elliptical galaxy.  Although this difference may
be explained in terms of a varying degree of star formation, the
similarities of their surface brightness profiles and the
submillimeter detection of 4C~39.37 might suggest that the intrinsic
spectra is equally blue (young stars or an AGN), and that the
difference is the amount of reddening.

\end{abstract}

\keywords{galaxies:active---galaxies:evolution---galaxies:individual(4C~40.36;\\
  4C~39.37$=$6C~1232$+$39)---radio continuum:galaxies}
\section{INTRODUCTION} 

Because of their high luminosities, radio galaxies are extremely
useful for studying the process of galaxy formation and evolution at
high redshift.  Although their substantial luminosities---the very
property which makes them so useful---may make them quite atypical of
galaxies, whatever is learned for high-redshift radio galaxies (HzRGs)
might also be applicable to a much broader spectrum of high-redshift
galaxies.

When using radio galaxies as a probe for galaxy formation/evolution at
high redshift, the important observations are of the near-IR continuum
emission.  Although the spectacular morphology of the luminous
line-emitting regions attracts our attention, the line-emitting
regions do not provide significant information on the underlying
stellar populations since they are most likely excited by the central
AGN \citep{Rawlings89,Willott99}.  On the other hand, the near-IR
continuum emission is thought to originate from the underlying stellar
population because of the remarkable tightness and continuity seen in
the $K$-band Hubble diagram ($K-z$ relation) of HzRGs as first shown
by \citet{Lilly84} (for the recent compilations of $K-z$
diagram, see \citet{Jarvis01} and \citet{Breuck02}).  Since the UV continuum is often
contaminated by scattered AGN light \citep[e.g.,][]{Vernet01}, the near-IR
continuum is the most robust observable from which we can hope to
extract information on the underlying stellar populations in HzRGs.

When deriving near-infrared continuum properties of HzRGs such as
magnitudes and morphology, it is essential to remove the contribution
from the luminous line-emitting regions \citep[e.g.,][]{Eales93}.
This becomes especially important when observing radio galaxies at $2
\lesssim z \lesssim 4$, where strong optical emission lines often
contaminate $H$- and $K$-band observations.  This redshift range is of
great interest since HzRGs start to show a dramatic evolution in terms
of morphology \citep{Breugel98} and submillimeter luminosity
\citep{Archibald01}, although the $K$-band (i.e., restframe visual)
luminosity evolution remains relatively mild \citep{Jarvis01}.

To study the properties of the near-infrared continuum sources in
HzRGs, we have been carrying out a near-infrared imaging and
spectroscopic study of HzRGs using the Keck, Hale, and Hubble Space
Telescopes (HST) (\citealt{Armus98}; \citealt{Egami99};
Armus~et~al.~2002, in preparation).  In this paper, we present observations
of two FR~II radio galaxies, 4C~40.36 at $z=2.27$ \citep{Chambers88}
and 4C~39.37 ($=$ 6C~1232$+$39) at $z=3.22$
\citep{Rawlings90,Eales93b}.  These galaxies were chosen for
HST/NICMOS observations because their restframe visual emission lines
(\oiii\ and \hanii\ for 4C~40.36 and \oiii\ for 4C~39.37) are
redshifted into the narrow-band filters available with HST/NICMOS.
Together with the high-quality ground-based data from the Keck and
Hale Telescopes, we are able to examine the nature of these galaxies
in detail.

To facilitate the comparison with previously published results, we
assume $\Omega_{M}=1$, $\Omega_{\Lambda}=0$, $H_{0}=50$ $h_{50}$ km
s$^{-1}$ Mpc$^{-1}$ throughout the paper.  With these parameters,
1\arcsec\ subtends 7.9, and 7.1 $h_{50}^{-1}$ kpc for 4C~40.36 and
4C~39.37, respectively.
\section{OBSERVATIONS AND DATA REDUCTION}

A log of all the observations is presented in Table~\ref{obs}, which
lists the telescope/instrument setups as well as observed wavelengths
and integration times.  Further details are described below.

\subsection{HST Near-IR Imaging}

The HST near-infrared images were taken with Camera~1 and Camera~2 of
the Near Infrared Camera and Multi-Object Spectrometer
\citep[NICMOS,][]{Thompson98}.  Both cameras use a 256$\times$256
HgCdTe array.  The pixel scale is 0\farcs043/pixel for Camera~1 and
0\farcs075/pixel for Camera~2.  The array was read out in the
MULTIACCUM mode with an exposure time of 512 seconds per image.

The images were reduced and calibrated with the NICMOS pipeline
calibration task CALNICA by running the IRAF command NICPIPE with
darks and flats taken with the same camera and readout parameters.
The bias drift between readouts and the quadrant-dependent sky/bias
level offsets were removed by using the IRAF command BIASEQ and
PEDSKY.  Once each image was reduced and calibrated, a Chebyshev
polynomial of order two was fitted and subtracted from each row and
column to remove the baseline slopes.  Finally, the DitherII package
\citep{Fruchter02} was used to remove cosmic rays and combine the
images.  Neither pixel scale change nor image rotation was applied
when producing the final dithered images, with which all the
photometry and surface brightness profile fitting were done.  Pixel
scale change and image rotation were applied only when producing
figures for visual display.

For the continuum flux calibration, the photometric calibrations of
$5.600\times10^{-6}$ (NIC1; F145M) and $3.760\times10^{-6}$
Jy~(DN/s)$^{-1}$ (NIC2; F187W) were adopted.  The emission line fluxes
were calculated from the narrow-band count rates according to the
prescription in the HST NICMOS Data Handbook with the photometric
calibrations of $6.148\times10^{-17}$ (NIC1; F164N),
$2.618\times10^{-17}$ (NIC2; F212N), and $2.693\times10^{-17}$
Jy~(DN/s)$^{-1}$ (NIC2; F215N).  The count rates were measured within a
4\arcsec-diameter beam.  

\subsection{Keck Near-IR Imaging and Spectroscopy}

The Keck near-infrared images were taken with the Near Infrared Camera
(NIRC; Matthews \& Soifer 1994) on the Keck I Telescope on Mauna Kea
in Hawaii.  NIRC uses a Hughes-SBRC 256$\times$256 InSb array, and is
attached to the f/25 forward Cassegrain focus of the telescope,
producing a pixel scale of 0\farcs15/pixel with a field of view of
38\arcsec\ on a side.

Low-resolution near-infrared spectra were also taken with NIRC.  The
slit width was set to 0\farcs68 (4.5 pixels), which gives a resolving
power (R=$\Delta \lambda$/$\lambda$) of $\sim 80$, and covers a
wavelength range $\delta \lambda$ of 1--2.4 $\mu$m in two wavelength
settings, $1.0-1.6$ $\mu$m and $1.4-2.5$ $\mu$m.  For 4C~40.36,
spectra were taken at these two settings while for 4C~39.37, only the
longer-wavelength setting was used.  The photometric calibration was
done by reference to the near-IR standard stars of \citet{Hawarden01}
and \citet{Hunt98}.

For spectroscopy, each target was first acquired by direct imaging
through a broad-band filter.  The slit was then aligned at a position
angle (east of north) of 83$\degr$ for 4C~40.36 and 140$\degr$ for
4C~39.37, respectively, which roughly correspond to the position angle
of the radio axis.  The object was moved by 5\arcsec\ to five
positions along the slit for successive integrations.

The data were processed in a conventional manner. The spectra were
corrected for atmospheric features by dividing the spectra by that of
a G star observed on the same night and at a similar air mass.  The
spectra were then multiplied by the blackbody spectrum with an
effective temperature corresponding to that of the G star.

The broad-band images show that in these two galaxies with a typical
seeing of 0.5\arcsec, the 0\farcs7 slit captures only 50\% of the
source flux measured in a 4\arcsec\ diameter beam because of the
significant spatial extent of the line-emitting regions.  Therefore,
the spectra were calibrated such that when summed over the passband of
the corresponding broad-band filter, they produced half of the flux
measured with a 4\arcsec\ diameter beam.

\subsection{Hale Near-IR Spectroscopy}

Moderate-resolution near-infrared spectra were obtained with the
long-slit near-infrared spectrograph \citep{Larkin96} on the Hale 5-m
Telescope at Palomar Observatory.  The slit width was set to 0\farcs7
(4 pixels), which gives a resolving power of $R\sim$ 1000 with a full
wavelength coverage $\delta \lambda$ of $\sim$ 0.1 $\mu$m.  The seeing
was estimated to be $\sim$ 0\farcs5 when the spectra were taken.

Each target was acquired by direct imaging through a 1 \% CVF filter
at the wavelength of the target line with an integration time of 300
seconds.  The slit was aligned at a position angle of 83$\degr$ for
4C~40.36 and 140$\degr$ for 4C~39.37.  The spectra were taken in
pairs, and the object was moved along the slit by 20\arcsec\ between
the two spectra.  After each pair, the object was moved by 5\arcsec\ in
one direction, and another pair was taken.

The spectra were corrected for atmospheric features by dividing the
spectra by that of a G star observed on the same night and at a
similar air mass.  The spectra were then multiplied by the blackbody
spectrum with an effective temperature corresponding to that of the G
star.  

Since the sky was not photometric when these spectra were taken, the
flux-calibration was done such that the line flux of the brightest
line in each spectrum has the value measured in the NIRC spectrum
since both the slit size and seeing were very similar.
\section{RESULTS}

\subsection{Morphology}

Figure~\ref{im4036} shows the HST and Keck images of 4C~40.36.  There
is a striking difference in morphology between the line-emitting
regions and the line-free continuum source.  The \hanii\ line-emitting
regions (Figure~\ref{im4036}a) and the \oiii\ line-emitting regions
(Figure~\ref{im4036}b) show a clumpy, nearly linear, filamentary
structure while the line-free continuum images taken at 1.45 and 1.87
$\mu$m (4430 \AA\ and 5720 \AA\ in the restframe) show a much more
compact and symmetric source (Figure~\ref{im4036}c).  A comparison
with the $K$-band image (Figure~\ref{im4036}d), which contains the
\hanii\ emission in the passband, clearly indicates that the line
emission dominates the $K$-band morphology of 4C~40.36.  The
line-emitting region extends over 2\arcsec\ ($\sim$ 16 $h_{50}^{-1}$
kpc) along a position angle of 111\arcdeg\ (east of north) with three
major knots.  The central knot coincides with the continuum source
while the other two knots have no detectable continuum emission.
Also, the eastern knot shows a significant north-south extension.

A major difference between these near-IR images and the visual images
published previously \citep{Chambers96} is that the former do not show
the strong alignment effect seen in the latter.  For example, there is
no detectable near-IR flux coming from the location of the western
radio peak (the cross to the right in Figure~\ref{im4036}), which is
seen as a conspicuous second peak in the visual.  Furthermore,
Figure~\ref{im4036}a and \ref{im4036}b clearly show that although both
the radio and line emission are extended roughly east-west, there is
a significant misalignment between the axes of the two components,
with a PA of 81$\degr$ and 111$\degr$, respectively.  The possibility
of such a misalignment was previously noted by \citet{Chambers96}.
Because of this, the relation between the radio jets and
restframe-visual emission lines is not clear.  This also means that
our spectra, for which the slit was aligned to the radio axis, will
miss a substantial amount of the extended line flux.

Figure~\ref{im3937} shows the HST and Keck images of 4C 39.37.  Again,
there is a significant difference in morphology between the \oiii\
line-emitting regions (Figure~\ref{im3937}a) and the line-free
continuum (Figure~\ref{im3937}b), although the difference is less
striking than that of 4C~40.36.  The \oiii\ line emitting region is
wedge-shaped while the line-free continuum (4430 \AA\ in the rest
frame) is compact and symmetric.  The line-free continuum source in
Figure~\ref{im3937}b shows a significant elongation along
PA~$=$~139$\degr$, which is almost the same as the PA of the radio
axis.

The continuum, though compact, is spatially extended in both galaxies,
as compared to the PSF defined by the stars seen to the north of 4C
40.36 in Figure~\ref{im4036}.  The surface brightness profiles were
computed based on the isophotal ellipse fitting using the ELLIPSE task
in STSDAS/IRAF.  The center position, ellipticity, and position angle
were all allowed to vary.  The measured profiles were then fitted with
an $R^{1/4}$-law profile and an exponential profile.  To find the
best-fit model, we subtracted PSF-convolved models from the HST/NICMOS
line-free continuum image, and minimized the residual.  Once the best
model was found, we measured its surface brightness profile by running
the ELLIPSE task with the same parameters as those used to measure the
profile in the HST/NICMOS image.

Figure~\ref{sb} shows the results of the surface brightness profile
fitting.  The profiles have a kink around 0\farcs4 from the center,
which is due to the similar structure seen in the PSF.  It can be seen
that both the exponential and $R^{1/4}$ laws produce acceptable fits.
The difficulty of distinguishing these two type of profiles was noted
previously with the Keck/NIRC images of HzRGs \citep{Breugel98}, and
our results show that this difficulty persists even with the
HST/NICMOS data to the depth of our images.  In the case of 4C~40.36,
the exponential-law fit requires a point-source core containing $\sim$
20 \% of the total continuum flux.  The best-fit parameters are listed
in Table~\ref{sbfit}.

\subsection{Photometry}

Table~\ref{cont} lists the photometric measurements derived from the
NIRC and the HST/NICMOS images with a 4\arcsec-diameter beam.  The
pure continuum flux density was measured to be $\sim$10 $\mu$Jy with
the three line-free HST filters (F145M and F187W for 4C~40.36; F187W
for 4C~39.37), and as shown in the table, the continuum flux accounts
only for 20--26\% of the broad-band flux measured with the Johnson
broad-band filters, whose passbands contain strong emission lines.

Table~\ref{cont} also shows that the line flux contributes only
7--25\%\ of the broad-band flux.  As a result, the sum of the
continuum and line fluxes falls substantially short (32--48\%) of what
is needed to account for the broad-band magnitudes.  As shown later,
this is due to the mismatch between the filter passbands and line
wavelengths/profiles.  Although a correction can be made based on the
observed line profiles in the spectra, the large loss of the line flux
due to the narrow slit width ($\sim 50$\%) makes such a correction
highly uncertain unless the velocity structure is exactly the same in
and out of the slit.  For this reason, we will not use the narrow-band
line images for photometry purposes.  This also implies that the
relative brightnesses of the line emitting regions seen in the
narrow-band images may be significantly affected by the underlying
velocity structures.

\subsection{Spectroscopy}
 
Figure~\ref{sp4036} shows the low-resolution Keck/NIRC spectrum of 4C
40.36.  The restframe visible spectrum of 4C 40.36 is dominated by
strong emission lines as seen in the figure.  The lines detected are
the classical strong lines seen in the spectra of radio galaxies,
\nev\ 3426 \AA, \oii\ 3727 \AA, \neiii\ 3869 \AA, \oiii\ 4959/5007
\AA, \oi\ 6300 \AA, \hanii\ 6563 \AA, and \sii\ 6716/6731 \AA.  There
is also a possible detection of the H$\beta$ line as a faint wing
blueward of the \oiii\ line.  A faint continuum is also clearly
detected in the spectrum.  The spatial extent of the continuum along
the slit is significantly smaller than that of the emission lines,
which is consistent with the imaging results.  The slope of the
continuum is found to be extremely flat in $f_{\nu}$ (the solid line
in Figure~\ref{sp4036}c).

Figure~\ref{sp3937} shows the NIRC spectrum of 4C~39.37.  It shows a
strong \oiii\ 4959/5007 line, and the line emission is significantly
more extended than the continuum.  Again, H$\beta$ line might be seen
just blueward of the \oiii\ line.  There is also a possible detection
of the \oii\ line, but this is again uncertain.  The continuum slope
is significantly redder than that of 4C~40.36 and flat in
$f_{\lambda}$ (i.e., $f_{\nu} \propto \nu^{-2}$).
Figure~\ref{sp3937}a also contains the spectrum of a nearby object to
north-west (see Figure~\ref{im3937}), which does not show any obvious
emission line.

Table~\ref{line} lists the line fluxes and continuum flux densities
measured in the spectra.  The continuum flux densities measured here
agree well with the values derived with the line-free filters
(Table~\ref{cont}), indicating that most of the continuum emission is
captured with the 0\farcs7 slit.  This is consistent with the observed
compact morphology of the continuum sources.  The line plus continuum
fluxes here also fall substantially short of that needed to account
for the broad-band magnitudes, and this is likely due to the extended
nature of the emission lines.  

Figures~\ref{sp_o3} and \ref{sp_ha} show the Palomar
moderate-resolution ($R \sim 1000$) spectra of 4C 40.36 around \oiii\
and \ha, respectively.  The complex spatial and spectral structures of
the lines indicate the multi-component nature of the line emitting
regions that is expected from the complex morphology seen in
Figure~\ref{im4036}.  The integrated \oiii\ line can be characterized
by two components: a narrower (FWHM $=$ 560 km s$^{-1}$) component at
the systemic velocity and a broader (FWHM $=$ 1670 km s$^{-1}$)
blueshifted ($\Delta v = -470$ km s$^{-1}$) component
(Figure~\ref{sp_o3}b).  The broader blueshifted component of \oiii\
emerges mainly from the central knot, which coincides with the
continuum source (Figure~\ref{sp_o3}c).  In the spectrum, a faint \hb\
line can also be seen.

To measure the fluxes of various lines, we assumed that each line
consists of the two components derived above with a relative flux
ratio of 4.9 as seen with the \oiii\ 5007 \AA\ line.  Other
constraints imposed were as follows: (1) In the spectrum shown in
Figure~\ref{sp_o3}, the wavelengths of the \hb\ and \oiii\ 4959 \AA\
lines were fixed with respect to that of \oiii\ 5007 \AA\ line.  Also,
a 1:3 ratio was assumed between \oiii\ 4959 \AA\ line and 5007 \AA\
lines; (2) In the spectrum shown in Figure~\ref{sp_ha}, the
wavelengths of the \nii\ and \sii\ lines were fixed with respect to
that of \ha.  Also, a 1:3 flux ratio was assumed between the \nii\ 6548
\AA\ and 6583 \AA\ lines.

In Figures~\ref{sp_o3}b and \ref{sp_ha}b, we also plot the
transmission curves of the NIC2 F164N and F215N filters.  From these
plots, it was estimated that these filters contain 29\% and 83\%, of
the \oiii\ and \hanii\ line flux, respectively.  

Figure~\ref{sp_o3b} shows the Palomar moderate-resolution spectrum of
4C~39.37 around \oiii.  Figure~\ref{sp_o3b}c shows that this galaxy
has two distinct line-emitting regions separated in space and
velocity.  The blue component, which is displaced 1\arcsec\ toward
$-\Delta X$ and $\sim$ -500 km/s blueshifted from the main component,
fits well with a Gaussian with a central wavelength of 2.107 $\mu$m
and a FWHM of 1230 km/s.  With this component fixed (flux was left to
vary), another Gaussian was added to fit the overall 5007 \AA\ line
profile, which resulted in another component 470 km/s redward with a
FWHM of 770 km/s.  The \oiii\ 4959 \AA\ line was fitted with two
Gaussians by assuming the same line-component structure and a 1:3
flux ratio with respect to the 5007 \AA\ line.

Again, the narrow-band filter (NIC2 F212N) misses a significant
fraction of the \oiii\ line flux (Figure~\ref{sp_o3b}b).  It was
estimated that 68 \% of the \oiii\ line flux falls outside the F212N
filter.

The line parameters derived in the moderate-resolution spectra are
listed in Table~\ref{line2}.

\section{DISCUSSION}

\subsection{Emission Lines}

\subsubsection{Total Flux}

As seen in Table~\ref{line}, the low-resolution spectra basically
recover the continuum flux densities similar to those measured in the
HST line-free images, but the line fluxes are too small to account for
the the observed broad-band fluxes.  This is expected since the
0\farcs7 slit has been found to miss 50\% of the broad-band fluxes,
most of which is in line emission.  About two thirds of the intrinsic
line flux is estimated to be missing from the spectra shown in
Figure~\ref{sp4036} and \ref{sp3937}, and when this correction is
made, we can recover all the broad-band flux.

Note that the slit-loss corrected line fluxes in Table~\ref{line} are
a factor of few larger than what we would derive from the narrow-band
observations (Table~\ref{cont}) by correcting for the passband
mismatch based on the line shapes seen in Figure~\ref{sp_o3}, \ref{sp_ha}, and
\ref{sp_o3b}.  This suggests that the line-emitting regions which fell
outside our slit have a velocity structure significantly different
from those seen in the long-slit spectra.  

\subsubsection{Contribution to the Broad-Band Magnitude}

Table~\ref{linecont} lists our estimates for the contribution of
emission lines to the broad-band magnitudes compared with those
determined by the previous studies.  The estimates can be made in two
ways, either from the observed continuum flux densities or from the
line fluxes.  The estimates based on the continuum fluxes consistently
give a line contribution of 70--80\% regardless of the observing modes
(imaging and spectroscopy), which lends high confidence to these
measurements.  On the other hand, the estimates based on the line
fluxes (we only discuss the spectroscopic results and disregard the
narrow-band imaging results) give numbers which are typically a factor
of few smaller than the continuum-based estimates, and this is likely
due to the loss of source flux falling outside the slit as we have
already discussed.  The two types of estimates become consistent with
each other when a slit-loss correction is made for the latter, which
was not applied in the previous studies.  Also, the significantly
lower signal-to-noise of the previous spectra may have resulted in
less accurate spectrophotometry.

\subsubsection{Line Ratios and Equivalent Widths} 

With our data, extinction can be explicitly determined for 4C~40.36
although the value is likely to vary significantly from position to
position.  The \ha/\hb\ ratio was measured to be 2.5, which is
consistent with the Case~B no-extinction value within the uncertainty.
This suggests very little internal extinction in this galaxy, which is
consistent with the extremely flat spectrum seen in the restframe
visual.  With no extinction correction applied, we find the following
values for the diagnostic line ratios for 4C~40.36:
\oiii (5007)/\hb $=$ 9.1, \nii (6583)/\ha $=$ 1.65, \oi/\ha $=$ 0.45,
and \sii/\ha $=$ 0.94. These line ratios clearly show that the
excitation level of the line-emitting regions is high and comparable
to that seen in Seyfert galaxies \citep{Veilleux87}.  This is
consistent with the previous studies by \citet{Iwamuro96},
\citet{Evans98}, and \citet{Carson01}.  Also, the \sii(6716)/\sii(6731) 
ratio of 0.4 indicates that the electron density of the ionized gas is
high ($> 10^{4}$ cm$^{-3}$).  For 4C~39.37, we find \oiii (5007)/\hb
$>$ 14, again indicating a Seyfert-like excitation.

The equivalent widths of the emission lines are extremely large.
Table~\ref{line} shows that the restframe equivalent widths measured
directly in the spectra are $> 1000$ \AA\ with the \oiii\ 4959/5007
and \hanii\ lines; if we correct for the slit loss, these values
become 3000--4000 \AA.  Though abnormally large, comparably large
equivalent-width lines have been observed in other HzRGs
\citep[e.g.,][]{Simpson99}.  In fact, if the line contribution to
broad-band magnitudes is $\sim 80$ \% as we derived, emission-line
equivalent widths must be this large.  It is difficult to produce such
large equivalent widths (e.g., W(\ha) $= 0.45 \times$ W(\hanii) $=$
1400 \AA\ in 4C~40.36) by a stellar population unless the population
is extremely young, a few Myrs at most based on the instantaneous
models in Starburst99 \citep{Leitherer99}.  The strong emission lines
of these HzRGs, together with the high excitation level derived from
the line ratios, are more likely to be excited by the central AGN.

\subsection{Continuum}

\subsubsection{Surface Brightness Profile and Luminosity}

Although slightly extended, the continuum sources in 4C~40.36 and
4C~39.37 are spatially compact.  If we adopt the $R^{1/4}$-law fits,
the effective radii ($r_{e}$) are 2.7 $h_{50}^{-1}$ kpc for 4C~40.36
and 1.8 $h_{50}^{-1}$ kpc for 4C~39.37, respectively.  This is a
factor of several smaller than those of $z \sim 1$ radio
galaxies, $r_{e} \sim 10$ $h_{50}^{-1}$ kpc, measured by
\citet{McLure00}.  A similar trend of HzRGs having a smaller effective
radius was also found by \citet{Pentericci01}, who reported that the
average effective radius of five HzRGs at $1.68 < z < 2.35$ is 5.4
$h_{50}^{-1}$ kpc, based on the measurements of five HzRGs with
HST/NICMOS at 1.6 $\mu$m.  The effective radii they derived are
somewhat larger than those we do but this difference may be due to the
fact that their filter passbands contain emission lines.  We also note
that two of their five HzRGs have very small effective radii,
1.6 and 1.7 $h_{50}^{-1}$ kpc, respectively.

It should be noted that in this discussion, we use the $R^{1/4}$ law
simply as a parametrization that allows us to compare the continuum
sizes of 4C~40.36 and 4C~39.37 with those of $z \sim 1$ radio
galaxies.  It is not implied here that the surface brightness profiles
of 4C~40.36 and 4C~39.37 actually follow the $R^{1/4}$ law.

4C~40.36 and 4C~39.37 both appear to have high surface brightnesses
compared to $z \sim 1$ radio galaxies.  This can be seen in
Figure~\ref{kormendy}, which plots the surface brightness at the
effective radius ($I_{e}$) versus effective radius ($r_{e}$) in the
restframe $B$ band.  This plot clearly shows that the restframe
$B$-band surface brightness of 4C~40.36 and 4C~39.37 are 1--1.5 mag
larger than what is expected from the $I_{e}-r_{e}$ correlation (often
referred to as Kormendy relation) of 3CR radio galaxies at $z \sim
0.8$ \citep{McLure00}.

The restframe $B$-band luminosities of the continuum sources in
4C~40.36 and 4C~39.37 turn out to be similar to those of $z \sim 1$
radio galaxies.  In the $R^{1/4}$-law parametrization, the total
luminosity is proportional to $r_{e}^2 I_{e}$, so the increase of the
surface brightness is offset by the decrease of effective radius.  The
line of a constant luminosity is shown in Figure~\ref{kormendy}.  This
results in a very modest, if any, luminosity evolution in the $K-z$
diagram for 4C~40.36 and 4C~39.37.  Using the flux density
measurements with the F145M and F187 filters, which roughly sample the
restframe $B$-band of 4C~40.36 and 4C~39.37, the absolute $B$
magnitudes were derived to be $-$23.3 and $-23.4+5 \log h_{50}$ mag,
respectively.  For comparison, the absolute $B$ magnitudes of $z \sim
0.8$ radio galaxies were also derived using the Cousins $I$ magnitudes
from \cite{McLure00}, which also measure the restframe $B$-band light.
The derived value, $M_{B} = (-23.0 \pm 0.4)+5 \log h_{50}$ mag,
indicates that the luminosities of 4C~40.36 and 4C~39.37 are
consistent with those of the average $z \sim 0.8$ 3CR radio galaxies.

\subsubsection{Spectra}

The restframe visual continuum spectra of these two HzRGs are very
different.  The continuum of 4C~40.36 is almost completely flat
($f_{\nu} = const.$, Figure~\ref{sp4036}) while the continuum of
4C~39.37 is as red as that of a present day giant elliptical galaxy
($f_{\nu} \propto \nu^{-2}$, Figure~\ref{sp3937}).  This difference is
clearly illustrated in Figure~\ref{lowspec}, in which the
low-resolution NIRC spectra shown in Figure~{\ref{sp4036} and
{\ref{sp3937}} are further rebinned to bring out the continuum shapes.

The flat restframe visible continuum of 4C~40.36 can be produced by
either an AGN or a young ($<$ 100 Mys) stellar population.
\citet{Vernet01} recently detected a significant amount of
polarization in the restframe UV continuum of a number of HzRGs,
including 4C~40.36, and reported that the scattering efficiency
appears almost independent of wavelength.  This suggests that the flat
restframe visible continuum of 4C~40.36 may be scattered AGN light.
On the other hand, there might be a sign of a small Balmer break seen
in the spectrum, and if it is real, it would suggest the stellar
origin of the flat continuum.  A comparison with a GISSEL96
instantaneous burst model\footnote{The Simple Stellar Population (SSP)
model of the Galaxy Isochrone Synthesis Spectral Evolution Library
(GISSEL96) by Bruzual and Charlot \citep{Bruzual93} was used.  The
model used here is the one with the solar metallicity, the Salpeter
IMF (0.1--125 M$_{\odot}$), and theoretical stellar spectra compiled
by \citet{Lejeune97} (bc96\_0p0200\_sp\_ssp\_kl96 in GISSEL).  An
instantaneous burst model was selected since it gives the fastest
evolution of the mass-to-light ratio, resulting in the maximum
estimate for the age and stellar mass.}  shows that the continuum
spectrum can be fitted by the light from a 50--100 Myr old population
with a mass of $3-5 \times 10^{10}$ M$_{\odot}$.  This mass, however
should be considered as a lower limit since the galaxy might contain a
separate population of old stars whose light is masked by that of the
young population, but contributes significantly to the total galaxy
mass.  Recently, the existence of such a young (100--200 Myr)
population was also inferred for two other HzRGs, 3C 256 at $z=1.82$
\citep{Simpson99} and 53W002 \citep{Motohara01}.

The red continuum of 4C~39.37 can be produced by either an old ($\sim$
1 Gyr) stellar population or a significant amount of dust reddening an
intrinsically much bluer source.  If the continuum light is of stellar
origin and suffers little reddening, the GISSEL96 instantaneous burst
model predicts roughly an age of 0.7--1.4 Gyrs and a mass of 2--3
$\times 10^{11}$ M$_{\odot}$.  On the other hand, the recent
submillimeter continuum detection of 4C~39.37 \citep{Archibald01}
might support the dust-reddening explanation.  The observed 850 $\mu$m
flux density ($3.86 \pm 0.72$ mJy) translates into to a restframe 200
$\mu$m luminosity ($\nu L\nu$) of $3 \times 10^{11}$ L$_{\sun}$,
implying that the far-infrared luminosity of 4C~39.37 is comparable to
those of local luminous IR galaxies (LIRGs, L$_{IR} > 10^{11}$
L$_{\sun}$).  The total infrared luminosity could be much larger if
the dust temperature is significantly higher than 20 K.  A large
amount of warm dust would then lend strength to the argument that
reddening, and not age, is responsible for the shape of the restframe
visual spectrum we measure in 4C~39.37.

The amount of reddening necessary to produce the continuum slope of
the 4C~39.37 spectrum from a flat spectrum like that of 4C~40.36
(whether it is scattered AGN light or young star light) is
$E(B-V)=0.5$ mag (Figure~\ref{lowspec}b), which results in a visual
magnitude extinction ($A_{\rm V}$) of 1.6 mag, using the extinction
law of \citet{Cardelli89}.  Such an amount of extinction could
significantly suppress the apparent luminosity evolution of HzRGs in
the $K-z$ diagram since the $K$ band samples the restframe visual for
these galaxies.

\subsection{Implications}

The small effective radius and high surface brightness of 4C~40.36 and
4C~39.37 (Figure~\ref{kormendy}) seem to indicate an increased level
of activity in the nuclear region of these HzRGs compared with $z \sim
1$ radio galaxies.  The nature of the luminosity source, however,
remains unknown since there is no clear detection of either stellar or
AGN spectral features in the restframe visual continua.  The surface
brightness profiles, which can be fitted by an $R^{1/4}$ law or an
exponential law, might be taken as the evidence that the restframe
visual light is of stellar origin, but recently there is a suggestion
that scattered AGN light may produce a similar surface brightness
profile (K.\ C.\ Chambers 2002, private communication).

Although our near-IR Keck/NIRC spectra of 4C~40.36 and 4C~39.37
provide vital information on the shape of the restframe visual
continuum, it is not possible to distinguish the various possibilities
listed above using our data alone.  For this, it is necessary to
determine the SED of the continuum sources in the restframe UV.  For
example, two hypotheses for the red continuum of 4C~39.37, an old
stellar population vs. dust reddening, are easily distinguishable in
the restrame UV because the former would show much less UV emission
than the latter based on our estimate on the reddening.  Although some
visual (restframe UV) data exist in the literature for 4C~40.36 and
4C~39.37, a direct comparison with the near-IR data presented here is
difficult because of their poor spatial resolution and strong
contamination from the component producing the alignment effect.

\section{SUMMARY AND CONCLUSIONS}

Using the Hubble, Keck, and Hale Telescopes, we have obtained
near-infrared images and spectra of two FR II HzRGs, 4C~40.36
($z=2.27$) and 4C~39.37 ($z=3.22$).  The main conclusions are as
follows: 
\begin{enumerate}
\item The contributions of the \oiii\ 5007 \AA\ and \hanii\ lines to
  near-IR broad-band magnitudes are found to be $\sim$ 80\%,
  substantially larger than the previous estimates of 20--40\%
  determined for the same galaxies.  

  \item The continuum sources in 4C~40.36 and 4C~39.37 are extremely
  compact ($r_{e} \sim 2-3$ kpc) and of high surface brightness
  ($I_{e}(B) \sim 18$ mag/$\sq$\arcsec) compared to $z \sim 1$ 3CR
  radio galaxies.  However, their restframe $B$-band luminosities are
  similar, showing very little luminosity evolution.

  \item The continuum spectral shapes of the two HzRGs in the
  restframe visual are very different: the continuum spectrum of
  4C~40.36 is almost completely flat ($f_{\nu}$=const) while the
  spectrum of 4C~39.37 is as red as that of a local gE galaxy.  The
  origin of the continuum emission is not yet clear since no obvious
  AGN feature (e.g., a broad line) or stellar feature (e.g., a
  continuum break) is seen.  

  \item The difference of the continuum spectral shapes may be
  explained in terms of a varying degree of star formation as well as
  different amounts of reddening.

\end{enumerate}

\acknowledgments

We thank Ken Chambers for making his results available to us before
publication, and Aronne Merrelli for comments on the manuscript.  The
authors wish to recognize and acknowledge the very significant
cultural role and reverence that the summit of Mauna Kea has always
had within the indigenous Hawaiian community.  We are most fortunate
to have the opportunity to conduct observations from this mountain.
Support for proposal \#7860 was provided by NASA through a grant from
the Space Telescope Science Institute, which is operated by the
Association of Universities for Research in Astronomy, Inc., under
NASA contract NAS~5-26555.

\clearpage

\begin{deluxetable}{llcclccr}
\scriptsize
\rotate
\tablewidth{0pt}
\tablecaption{Observing Logs \label{obs}}
\tablehead{
\colhead{Object} & \colhead{UT date} & \colhead {Telescope} &
\colhead{Instrument} & \colhead{Filter} & \colhead{$\lambda$} &
\colhead{Line} & \colhead{Exp time} \\
\colhead{} & \colhead{} & \colhead{} & 
\colhead{} & \colhead{} & \colhead{($\mu$m)} &
\colhead{} & \colhead{(sec)}
}
\startdata
4C~40.36 & 1996 September 4 & Keck I & NIRC  & $K$      & 2.01 -- 2.44   &           & 1080 \\
         &             &        &       & \feii\   & 1.638 -- 1.656 & \oiii\    & 1680 \\
         &             &        &       & Spectrum & 1.0 -- 1.6     &           & 2000 \\
         &             &        &       & Spectrum & 1.4 -- 2.5     &           & 2000 \\
         & 1996 August  1 & Hale   & NSPEC & Spectrum & 2.11 -- 2.21   &           & 3600 \\
         &             &        &       & Spectrum & 1.57 -- 1.66   &           & 3600 \\
         & 1998 September 18 & HST    & NIC1  & F164N    & 1.637 -- 1.655 & \oiii\    & 7168 \\
         &             &        &       & F145M    & 1.35 -- 1.55   & continuum & 2560 \\
         & 1998 September 24 & HST    & NIC2  & F215N    & 2.139 -- 2.159 & \hanii\   & 7168 \\
         &             &        &       & F187W    & 1.75 -- 2.00   & continuum & 2560 \\ 
4C~39.37 & 1996 June  1 & Keck I & NIRC  & $K$      & 2.008 -- 2.435 &           &  900 \\
         &             &        &       & Spectrum & 1.4--2.5       &           & 4000 \\
         & 1997 February 22 & Hale   & NSPEC & Spectrum & 2.11 -- 2.21   &           & 3600 \\
         & 1998 August  1 & HST    & NIC2  & F212N    & 2.111 -- 2.132 & \oiii\    & 7168 \\
         &             &        &       & F187W    & 1.75 -- 2.00   & continuum & 3072 \\
\enddata
\end{deluxetable}

\clearpage

\begin{deluxetable}{lcccccccccccc}
\tablecolumns{13}
\tablewidth{0pt}
\rotate
\tablecaption{Surface Brightness Profile Fit in the F187W images  \label{sbfit}}
\tablehead{
\colhead{Object} & \colhead{} & \multicolumn{5}{c}{$R^{1/4}$ law} & \colhead{} &
\multicolumn{5}{c}{Exponential law} \\
\cline{3-7} \cline{9-13} 
\colhead{} & \colhead{} &
\multicolumn{2}{c}{$r_{e}$} & \colhead{} &\multicolumn{2}{c}{$I_{e}$} & \colhead{} &
\multicolumn{2}{c}{$r_{d}$} & \colhead{} &\multicolumn{2}{c}{$I_{d}$} \\
\colhead{} & \colhead{} &
\colhead{(\arcsec)} & \colhead{($h_{50}^{-1}$ kpc)} & \colhead{} &
\colhead{($\mu$Jy/$\sq$\arcsec)} & \colhead{($B$ mag/$\sq$\arcsec)\tablenotemark{a}} & \colhead{} &
\colhead{(\arcsec)} & \colhead{($h_{50}^{-1}$ kpc)} & \colhead{} & 
\colhead{($\mu$Jy/$\sq$\arcsec)} & \colhead{($B$ mag/$\sq$\arcsec)\tablenotemark{a}} 
}
\startdata
4C $+$40.36 & & 0.34 & 2.7 & & 5.0 & 18.4 & & 0.19 & 1.5 & &  33 & 16.4 \\
4C $+$39.37 & & 0.26 & 1.8 & & 5.9 & 17.5 & & 0.14 & 1.0 & &  58 & 15.0 \\
\enddata

\tablecomments{$R^{1/4}$-law profile: $I(r) = I_{e} \exp\{-7.67[(r/r_{e})^{0.25} -1]\}$; 
Exponential profile: $I(r) = I_{d} \exp (-r/r_{d})$}
\tablenotetext{a}{
In the calculation of the restframe $B$-band surface brightnesses,
only the $(1+z)^{3}$ cosmological dimming (in $f_{\nu}$) was taken
into account.  This is justified because, (1) for 4C~39.37, the F187W
filter passband roughly samples the restframe $B$-band light at
$z=3.2$, and (2) for 4C~40.36, our NIRC spectra show that its
continuum is almost completely flat in $f_{\nu}$.}

\end{deluxetable}

\clearpage

\begin{deluxetable}{cccccccc}
\tabletypesize{\footnotesize}
\rotate
\tablewidth{0pt}
\tablecaption{Line/continuum measurements with the NIRC HST/NICMOS images\label{cont}}
\tablehead{
\colhead{Object} & \multicolumn{2}{c}{Filter} & \colhead{Instrument} &
\multicolumn{2}{c}{Continuum} & \multicolumn{2}{c}{Line} \\
\colhead{}     & \colhead{Name} & \colhead{Line}  & \colhead{} &
\colhead{} & \colhead{} & \colhead{} & \colhead{} \\
\colhead{}   & \colhead{} & \colhead{} & \colhead{} &
\colhead{($\mu$Jy)} & \colhead{(\%)\tablenotemark{a}} &
\colhead{(10$^{-18}$ W m$^{-2}$)} & \colhead{(\%)\tablenotemark{b}}}
\startdata
4C $+$40.36 & F145M &         & NIC1 & $10 \pm 2$ &  20 &                &      \\
            & F164N & \oiii\  & NIC1 &            &     &  $1.9 \pm 0.3$ & 12   \\
            & H     &         & NIRC & $50 \pm 1$ & 100 &                &      \\
            &       &         &      &            &     &                &      \\
            & F187W &         & NIC2 & $12 \pm 1$ &  23 &                &      \\
            & F215N & \hanii\ & NIC2 &            &     &  $3.4 \pm 0.1$ & 25   \\
            & K     &         & NIRC & $53 \pm 1$ & 100 &                &      \\
            &       &         &      &            &     &                &      \\
4C $+$39.37 & F187W &         & NIC2 & $6 \pm 1 $ &  26 &                &      \\
            & F212N & \oiii\  & NIC2 &            &     &  $0.6 \pm 0.1$ & 7    \\
            &   K   &         & NIRC & $32 \pm 1$ & 100 &                &     
\enddata

\tablenotetext{a}{ The percentage contribution of the continuum flux
  to the total flux as measured in the corresponding broad-band
  filters.  The measured flux densities were adjusted to the central
  wavelength of the corresponding Johnson filter by assuming a
  continuum shape of $f_{\nu} \propto \nu^{0}$ for 4C~40.36 and
  $f_{\nu} \propto \nu^{-2}$ for 4C~39.37.  }

\tablenotetext{b}{
The percentage contribution of the line
flux to the total flux as measured in the corresponding broad-band
filters.
}

\end{deluxetable}

\clearpage

\begin{deluxetable}{llccccrcrrcrr}
\rotate
\tabletypesize{\footnotesize}
\tablewidth{0pt}
\tablecaption{Line/continuum measurements with the low-resolution spectra \label{line}}
\tablehead{
\colhead{Object} & \colhead{} & \colhead{$\lambda_{rest}$} & 
\colhead{$\lambda_{obs}$} & $z$ & \multicolumn{2}{c}{Continuum} & \multicolumn{3}{c}{Line} & 
\multicolumn{3}{c}{Corrected Line\tablenotemark{a}}\\
\colhead{} & \colhead{}    & \colhead{} & \colhead{} & \colhead{} & \colhead{} &
\colhead{} & \colhead{Flux} & \colhead{} & \colhead{EW} & \colhead{Flux} & \colhead{} & \colhead{EW} \\
\colhead{} & \colhead{}    & \colhead{(\AA)} & \colhead{($\mu$m)} & 
\colhead{} & \colhead{($\mu$Jy)} & \colhead{(\%)\tablenotemark{b}} & 
\colhead{($10^{-18}$ W m$^{-2}$)} & \colhead{(\%)\tablenotemark{c}} & \colhead{(\AA)} &
\colhead{($10^{-18}$ W m$^{-2}$)} & \colhead{(\%)\tablenotemark{c}} & \colhead{(\AA)}
}
\startdata
4C $+$40.36 & \nev\    & 3426            & 1.127   & 2.29    &              &    & 0.42$\pm$0.11 &    &   94 & 1.5  &    & 329  \\
            & \oii\    & 3727            & 1.219   & 2.27    &              &    & 1.37$\pm$0.13 &    &  306 & 4.8  &    & 1071 \\
            & \neiii   & 3869            & 1.270   & 2.28    &              &    & 0.30$\pm$0.13 &    &   67 & 1.1  &    & 235  \\
            & $J$      &                 &         &         &  7.3$\pm$0.6 &    &               &    &      &      &    &      \\
            & \hb\     & 4861            & \nodata & \nodata &              &    & $<$ 0.16      &    &      &      &    &      \\
            & \oiii\   & 4959/5007       & 1.638   & 2.27    &              &    & 4.96$\pm$0.09 & 31 & 1458 & 14   & 87 & 4084 \\
            & $H$      &                 &         &         &  9.1$\pm$0.8 & 18 &               &    &      &      &    &      \\
            & \oi\     & 6300            & 2.060   & 2.27    &              &    & 0.45$\pm$0.09 &  3 &  166 & 1.2  & 9  & 432  \\
            & \hanii   & 6563, 6548/6583 & 2.143   & 2.27    &              &    & 3.24$\pm$0.09 & 24 & 1194 & 8.4  & 61 & 3104 \\
            & \sii\    & 6716/6731       & 2.194   & 2.26    &              &    & 0.95$\pm$0.09 &  7 &  350 & 2.5  & 18 & 910  \\
            & $K$      &                 &         &         & 13.5$\pm$0.8 & 25 &               &    &      &      &    &      \\
4C $+$39.37 & \hb      & 4861            & \nodata & \nodata &              &    & $<$ 0.1       &    &      &      &    &      \\
            & \oiii\   & 4959/5007       & 2.101   & 3.20    &              &    & 1.90$\pm$0.03 & 23 & 1210 & 5.9  & 71 & 3751 \\
            & $K$      &                 &         &         &  7.8$\pm$0.5 & 24 &               &    &      &      &    &      \\
\enddata

\tablenotetext{a}{The slit-loss correction factors are 3.5 (4C~40.36:
            \nev, \oii, \neiii), 2.8 (4C~40.36: \oiii), 2.6 (4C~40.36:
            \oi, \hanii, \sii), and 3.1 (4C~39.37: \oiii).  These
            factors were derived by assuming that the slit captures
            all the continuum flux but only 50 \% of the
            total flux measured within a 4''-diameter beam.}
\tablenotetext{b}{The percentage contribution of continuum emission to the corresponding
            broad-band flux ($H$ or $K$).}
\tablenotetext{c}{The percentage contribution of line emission to the corresponding
            broad-band flux ($H$ or $K$).}

\end{deluxetable}

\clearpage

\begin{deluxetable}{llccrc}
\tablewidth{0pt}
\tabletypesize{\small}
\tablecaption{Line measurements with the moderate-resolution spectra \label{line2}}
\tablehead{
\colhead{Object} & \colhead{Line} & \colhead{$\lambda_{rest}$} &
\colhead{$z$} & \colhead{FWHM} & \colhead{Observed flux} \\
\colhead{} & \colhead{}    & \colhead{\AA} & 
\colhead{} & \colhead{(km s$^{-1}$)} & \colhead{($10^{-18}$ W m$^{-2}$)} 
}
\startdata
4C $+$40.36 & \hb\   & 4861 & 2.266 & 1670 & 0.34 \\
            &        &      & 2.272 &  560 & 0.07 \\
            & \oiii\ & 4959 & 2.266 & 1670 & 1.03 \\
            &        &      & 2.272 &  560 & 0.21 \\     
            & \oiii\ & 5007 & 2.266 & 1670 & 3.10 \\
            &        &      & 2.272 &  560 & 0.62 \\
            & \nii\  & 6548 & 2.266 & 1670 & 0.46 \\
            &        &      & 2.272 &  560 & 0.09 \\
            & \ha\   & 6563 & 2.266 & 1670 & 0.84 \\
            &        &      & 2.272 &  560 & 0.17 \\
            & \nii\  & 6583 & 2.266 & 1670 & 1.39 \\
            &        &      & 2.272 &  560 & 0.28 \\
            & \sii\  & 6716 & 2.266 & 1670 & 0.20 \\
            &        &      & 2.272 &  560 & 0.04 \\
            & \sii\  & 6731 & 2.266 & 1670 & 0.46 \\
            &        &      & 2.272 &  560 & 0.09 \\
4C $+$39.37 & \oiii\ & 4959 & 3.208 & 1230 & 0.19 \\
            &        &      & 3.214 &  770 & 0.27 \\  
            &        & 5007 & 3.208 & 1230 & 0.58 \\
            &        &      & 3.214 &  770 & 0.82 \\
\enddata
\tablecomments{Because of the various constraints applied to the
            spectral line fitting, many of the numbers listed here are
            not independent.  Wavelengths (i.e., redshifts), FWHM
            widths, and fluxes of some emission lines are fixed with
            respect to each other.  See text for more details on the
            fitting constraints.}
\end{deluxetable}

\clearpage

\begin{deluxetable}{lccccccccccc}
\rotate
\tablewidth{0pt}
\tablecaption{Line Contribution Estimates\tablenotemark{a} \label{linecont}}
\tablehead{
\colhead{} & \colhead{} & \multicolumn{3}{c}{From Continuum} & \colhead{} & \multicolumn{6}{c}{From Line} \\
\cline{3-5} \cline{7-12} \\
\colhead{Object} & \colhead{Band} & \multicolumn{2}{c}{This Work} & \colhead{I96} 
& \colhead{} & \multicolumn{2}{c}{This Work} & \colhead{ER96} &
\colhead{I96} & \colhead{E98} & \colhead{C01}\\
\colhead{} & \colhead{} & \colhead{Image} & \colhead{Spectra} & \colhead{} & \colhead{} &
\colhead{Spectra} & \colhead{Slit-loss corrected} & \colhead{} & \colhead{} & \colhead{} & \colhead{}\\
\colhead{} & \colhead{} & \colhead{(\%)} & \colhead{(\%)} & \colhead{(\%)} & \colhead{} &
\colhead{(\%)} & \colhead{(\%)} & \colhead{(\%)} & \colhead{(\%)} & \colhead{(\%)} & \colhead{(\%)}
}
\startdata
4C $+$40.36 & H & 80 & 82 & 77      &  & 31 & 87  & \nodata & 33      & \nodata & 81 \\
            & K & 77 & 75 & \nodata &  & 34 & 88  &     25  & \nodata & 44      & 54 \\
4C $+$39.37 & K & 74 & 76 & \nodata &  & 23 & 71  & 30      & \nodata & \nodata & \nodata \\   
\enddata

\tablenotetext{a}{The percentage contribution of the line
flux to the total flux as measured in the corresponding broad-band
filter.}

\tablerefs{
(1) ER96: \citealt{Eales96}; (2) I96: \citealt{Iwamuro96}; (3) E98:
\citealt{Evans98}; (4) C01: \citealt{Carson01}
}

\end{deluxetable}

\clearpage

\begin{figure}

  \vspace*{-1in}

  \hspace*{1cm}\psfig{file=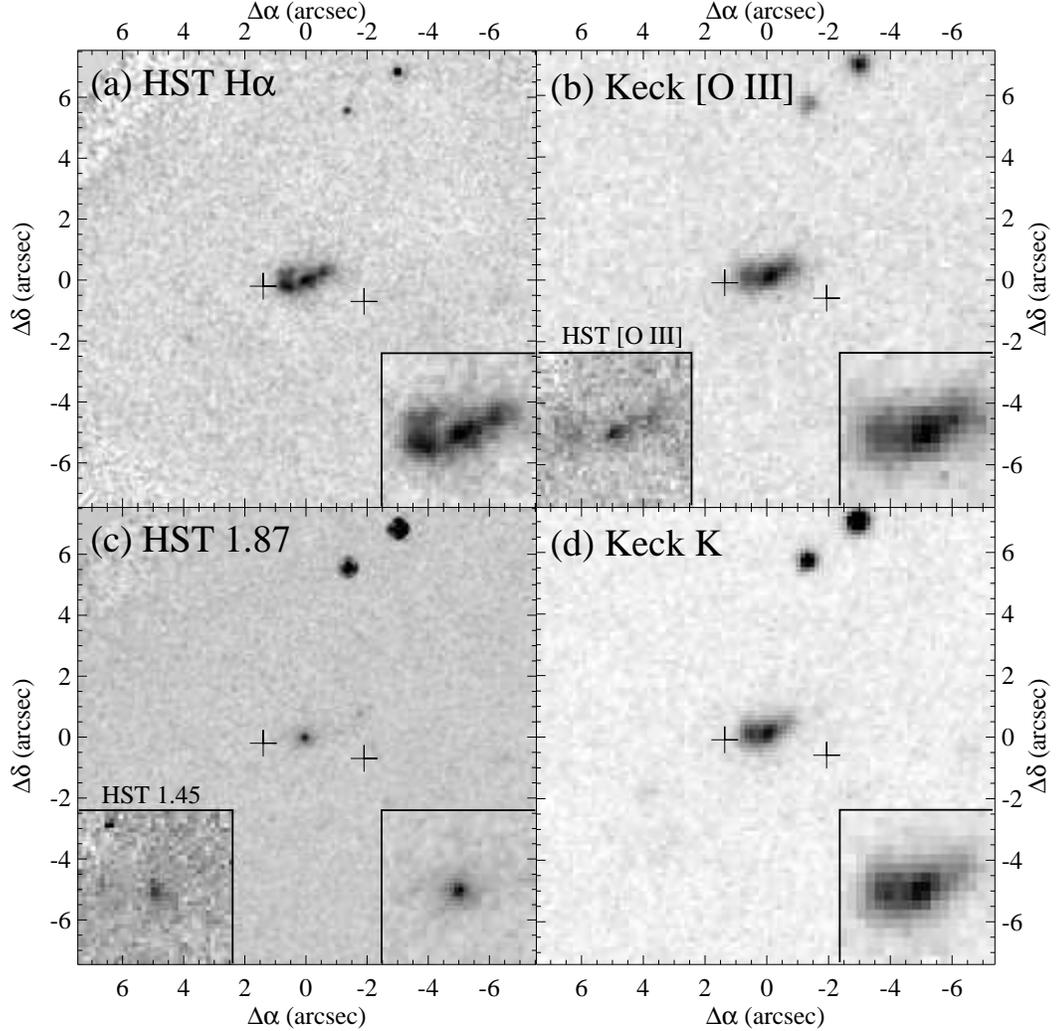,width=6in}

  \vspace*{-1in}

  \figcaption[fig1.ps]{Near-IR images of 4C~40.36 (north is up and
  east is left).  In each panel, the insets are $\times$2 magnified
  images of the 2\farcs55$\times$2\farcs55 around the radio galaxy:
  (a) the \hanii\ image taken with HST/NIC2 at 2.15$\mu$m; (b) the
  \oiii\ images taken with NIRC/Keck at 1.65 $\mu$m and with HST/NIC1
  at 1.64 $\mu$m (lower-left inset); (c) the line-free continuum
  images taken with HST/NIC2 at 1.87$\mu$m and with HST/NIC1 at
  1.45$\mu$m (lower-left inset); (d) Keck/NIRC $K$-band image.  The
  NIC1 images were resampled to the 0\farcs075/pixel pixel scale to
  reduce the noise.  The crosses indicate the radio lobe positions,
  and the registration was done by using the 4710 MHz map of
  \citet{Carilli97} with the assumption that the central faint radio
  peak coincides with the near-IR continuum source (the radio axis PA $=$
  81$\degr$).  Two field stars are also seen to the north.
  \label{im4036}}

\end{figure}

\begin{figure}

  \vspace*{-1in}

  \hspace*{1cm}\psfig{file=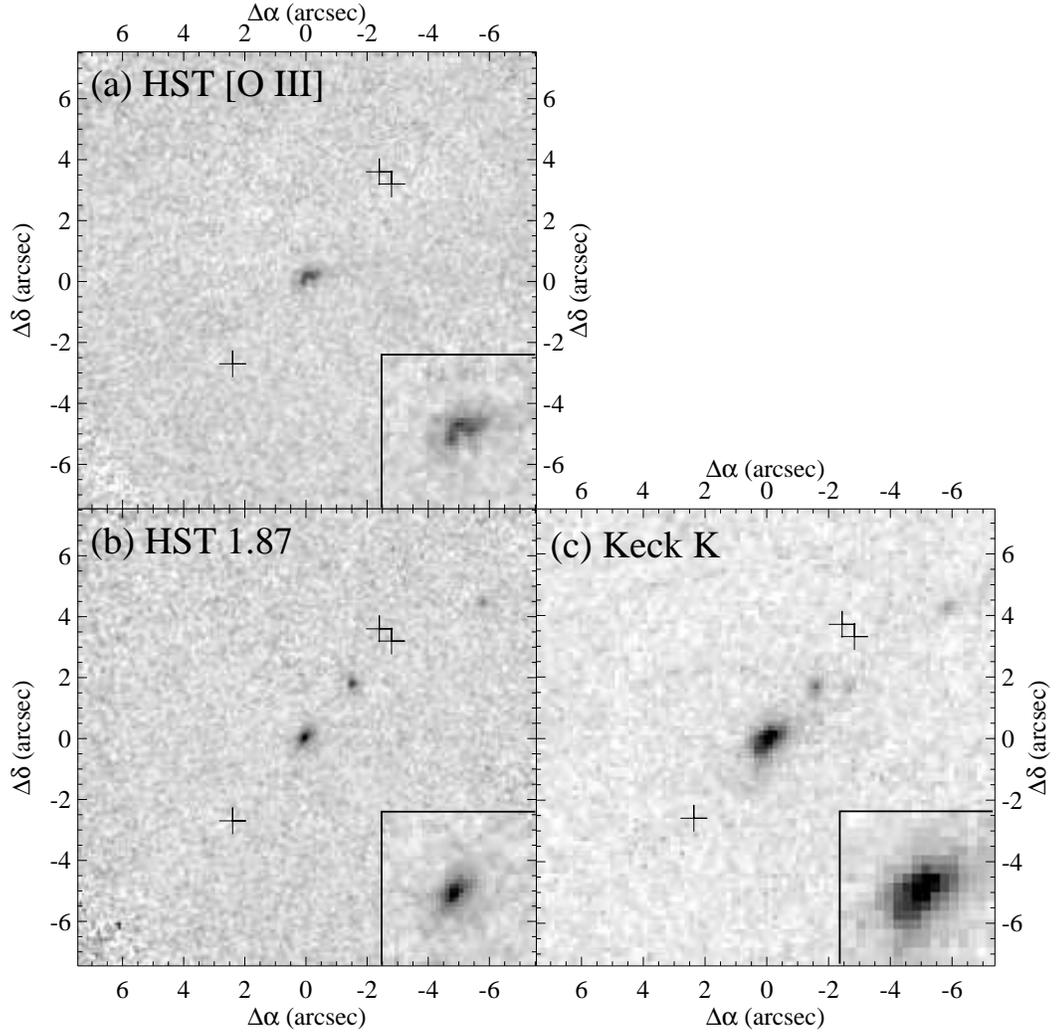,width=6in}

  \vspace*{-1in}

  \figcaption[fig2.ps]{Near-IR images of 4C~39.37 (north is up and
  east is left).  In each panel, the insets are $\times$2 magnified
  images of the 2\farcs55$\times$2\farcs55 around the radio galaxy:
  (a) the \oiii\ image taken with HST/NIC2 at 2.12 $\mu$m; (b) the
  line-free continuum image taken with HST/NIC2 at 1.87$\mu$m; (c)
  Keck/NIRC $K$-band image.  The crosses indicate the radio lobe
  positions, and the registration was done by using the 8210 MHz map
  of \citet{Carilli97} with the assumption that the central faint
  radio peak coincides the near-IR continuum source. The north-west
  lobe has two radio peaks (the radio axis PA $=$ 139,
  143$\degr$). \label{im3937}}

\end{figure}

\begin{figure}

  \hspace*{1cm}\psfig{file=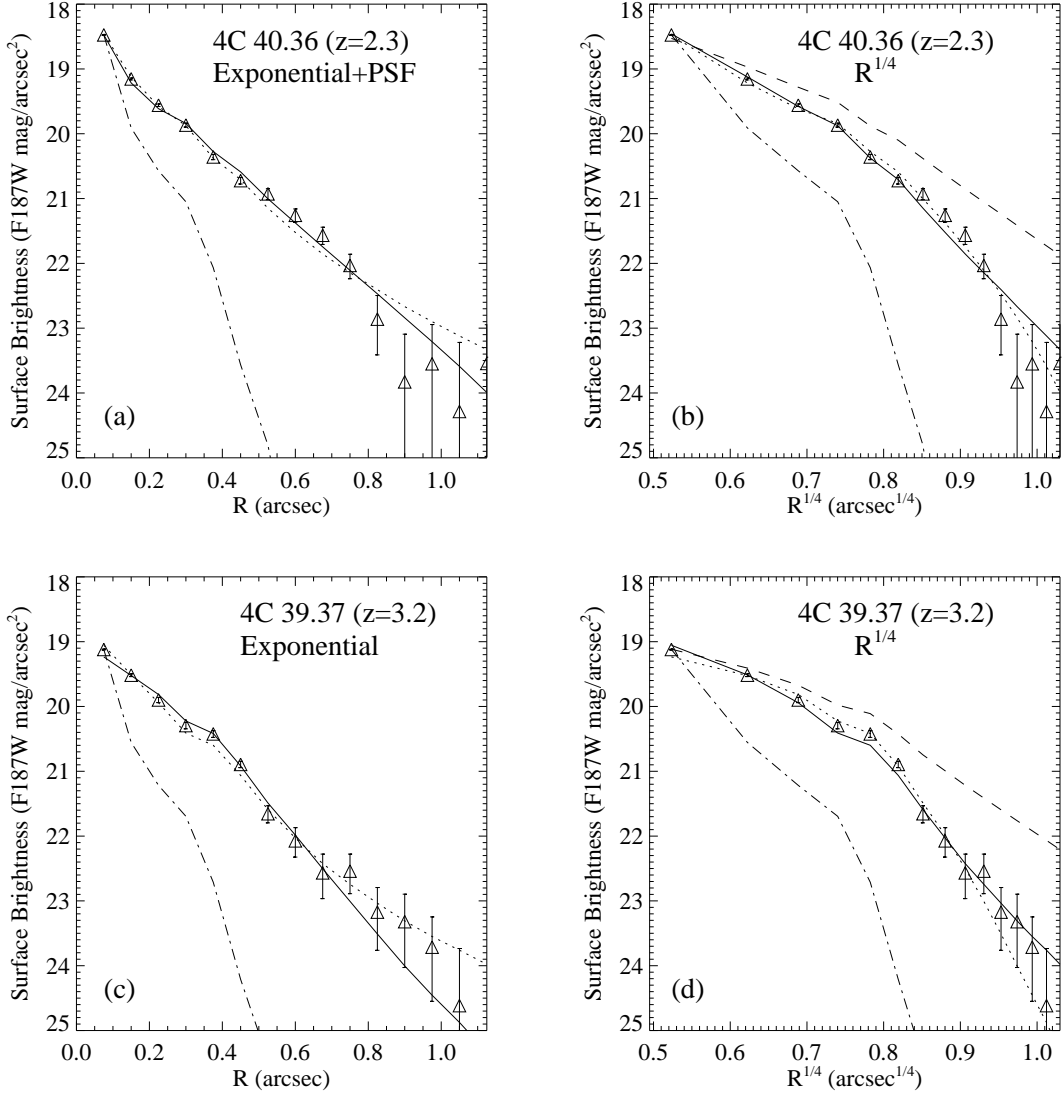,width=6in}

  \figcaption[fig3.ps]{The
  surface brightness profile fits to the line-free continuum sources
  seen in the HST/NIC2 1.87 $\mu$m images.  The surface brightness is
  expressed in the Vega-normalized F187W magnitude.  For each object,
  both the exponential-law fit (a and c) and the $R^{1/4}$-law fit (b
  and d) were performed.  In each panel, the dash-dot line indicates
  the PSF measured from the field star seen in Figure~\ref{im4036}
  while the solid line indicates the fit with the functional form
  identified in the panel.  The dotted line shows the fit with the
  other functional form (i.e., the $R^{1/4}$ fit in (a) and (c) and
  the exponential fit in (b) and (d)).  The dashed line in (b) and (d)
  shows the $R^{1/4}$-law profile with $r_{e}=$ 10 kpc for comparison.
  In (a), a PSF with a 20\% flux was added to fit the central peak.
  \label{sb}}

\end{figure}

\begin{figure}

  \vspace*{-0.5in}

  \hspace*{0.5in}\psfig{file=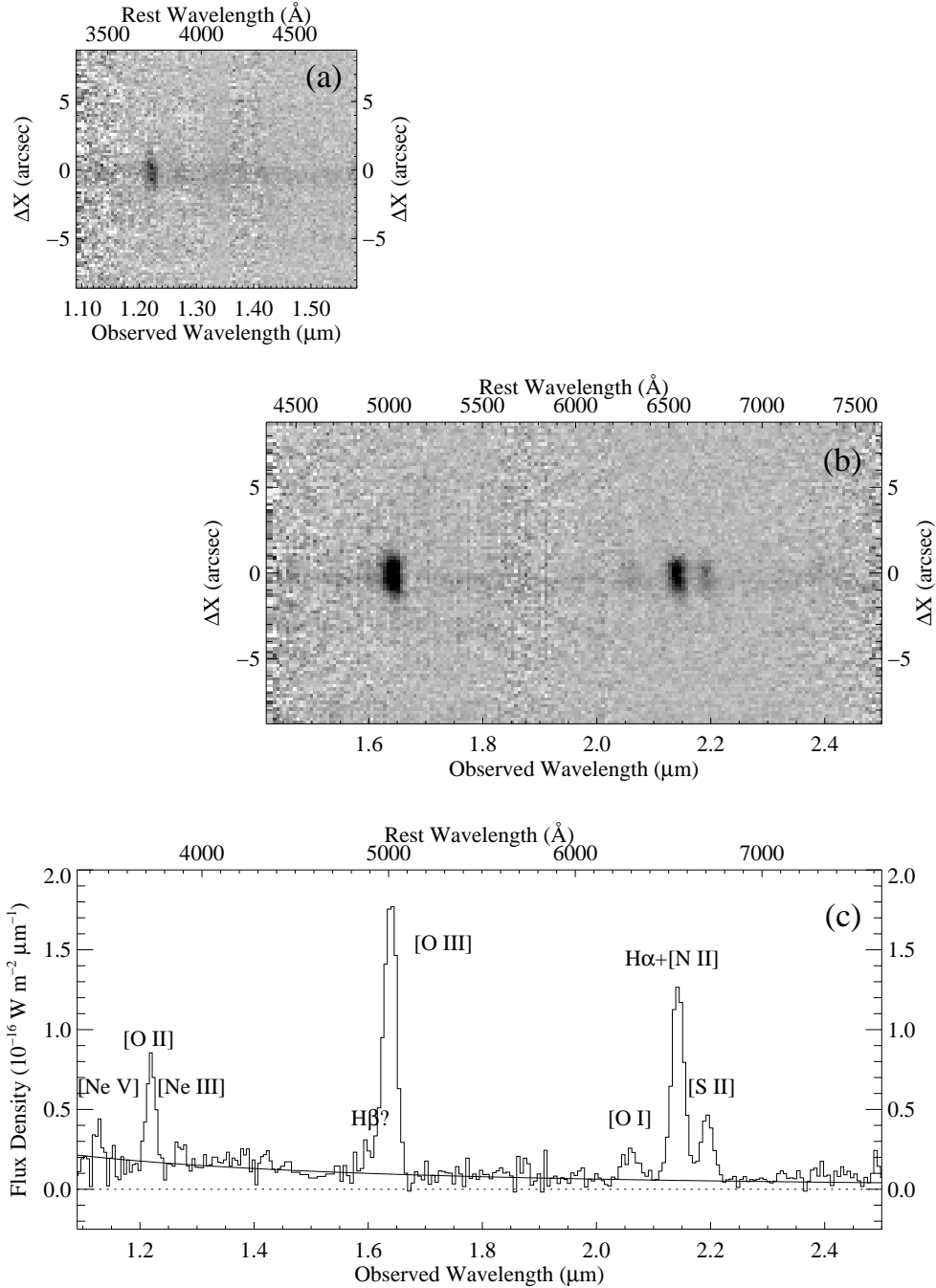,width=5.5in}

  \figcaption[fig4.ps]{The Keck/NIRC near-IR spectrum of 4C 40.36 with
  (a) the 1.0--1.6 $\mu$m grism and (b) the 1.4--2.5 $\mu$m grism.
  The $-\Delta$X direction corresponds to a position angle of
  83$\degr$ (E of N). A composite one-dimensional spectrum is shown in
  (c).  The solid line in (c) shows a flat spectrum in $f_{\nu}$,
  which provides a good fit to the continuum spectrum.
  \label{sp4036}}

\end{figure}

\begin{figure}

  \hspace*{-0.5cm}\psfig{file=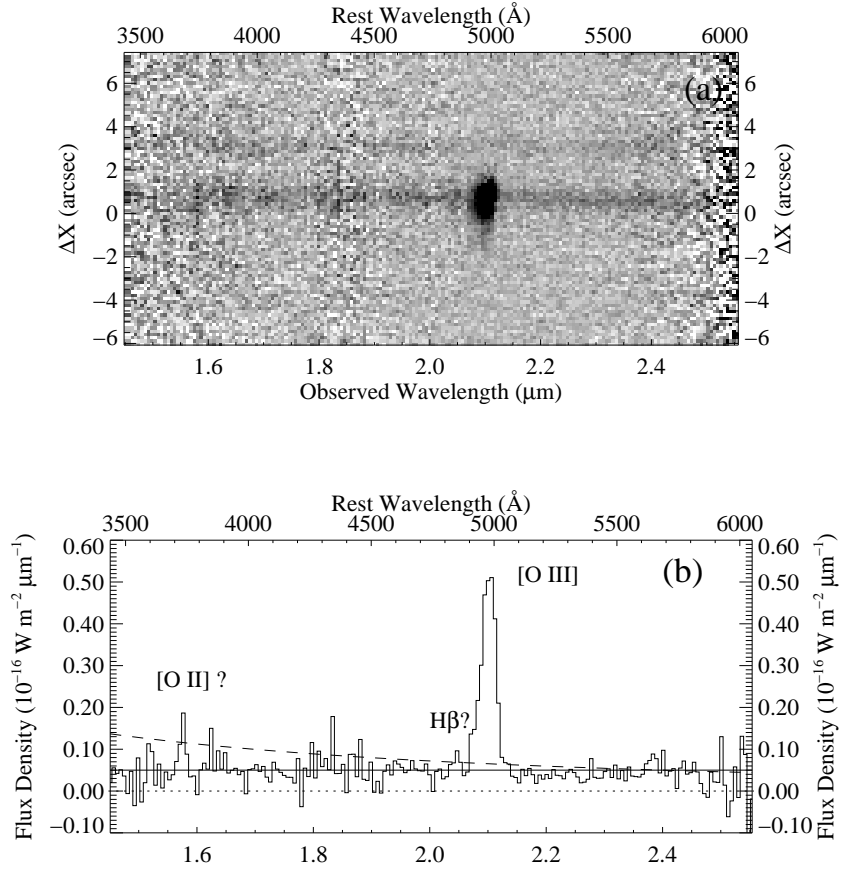,width=7in,angle=90}

  \figcaption[fig5.ps]{The Keck/NIRC near-IR spectrum of 4C~39.37 with
  the 1.4--2.5 $\mu$m grism.  In (a), the $-\Delta$X direction
  corresponds to a position angle of 140$\degr$ (E of N).  The lower
  spectrum is that of the radio galaxy while the upper one is that of
  a nearby object in the north west seen in the upper right corner of
  Figure~\ref{im3937}b and \ref{im3937}c.  A one-dimensional spectrum
  is shown in (b).  The solid line indicates a $f_{\nu} \propto
  \nu^{-2}$ spectrum, which is a good fit to the continuum spectrum
  of 4C~39.37, while the dashed line shows a $f_{\nu} = const.$
  spectrum (i.e., the continuum spectrum of 4C~40.36) for comparison.
  \label{sp3937}}

\end{figure}

\begin{figure}

  \vspace*{-1cm}

  \hspace*{1cm}\psfig{file=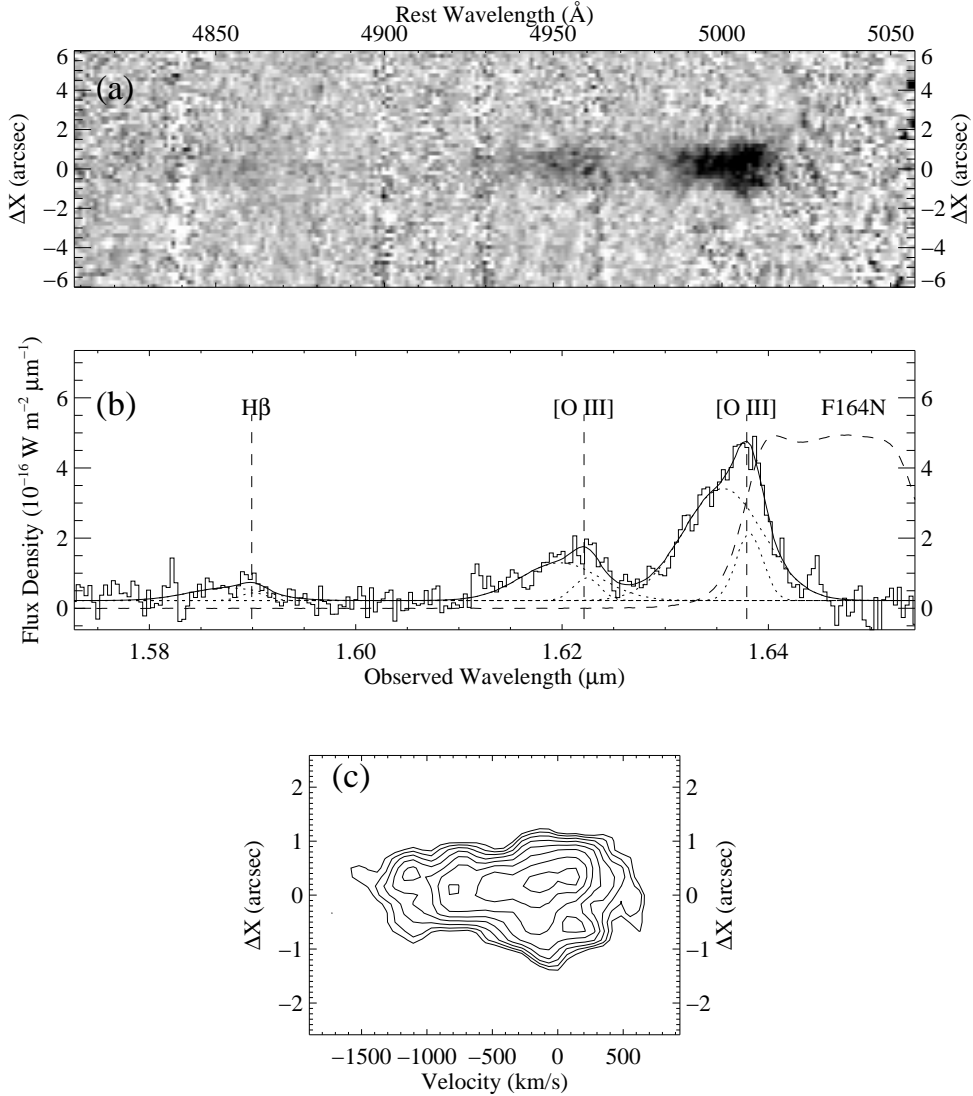,width=5.5in}

  \vspace*{-2cm}
  
  \figcaption[fig6.ps]{The \oiii\ spectrum of 4C~40.36 taken with the
  Palomar long-slit near-IR spectrograph: (a) The two dimensional
  image of the spectrum; (b) The summed one-dimensional spectrum.
  There is a possible detection of a faint H$\beta$ line.  The solid
  line shows our model fit with two velocity components shown by the
  dotted lines (see the text).  The dashed curve is the transmission
  of the HST/NIC1 F164N filter; (c) The intensity contours of the
  \oiii\ 5007\AA\ line as a function of position and line-of-sight
  velocity.  The spectral image (a) was smoothed with a 2-pix FWHM
  Gaussian.  In the panels (a) and (c), the $-\Delta$X direction
  corresponds to a position angle of 84$\degr$ (E of
  N). \label{sp_o3}}

\end{figure}

\begin{figure}

  \vspace*{-1cm}

  \hspace*{1cm}\psfig{file=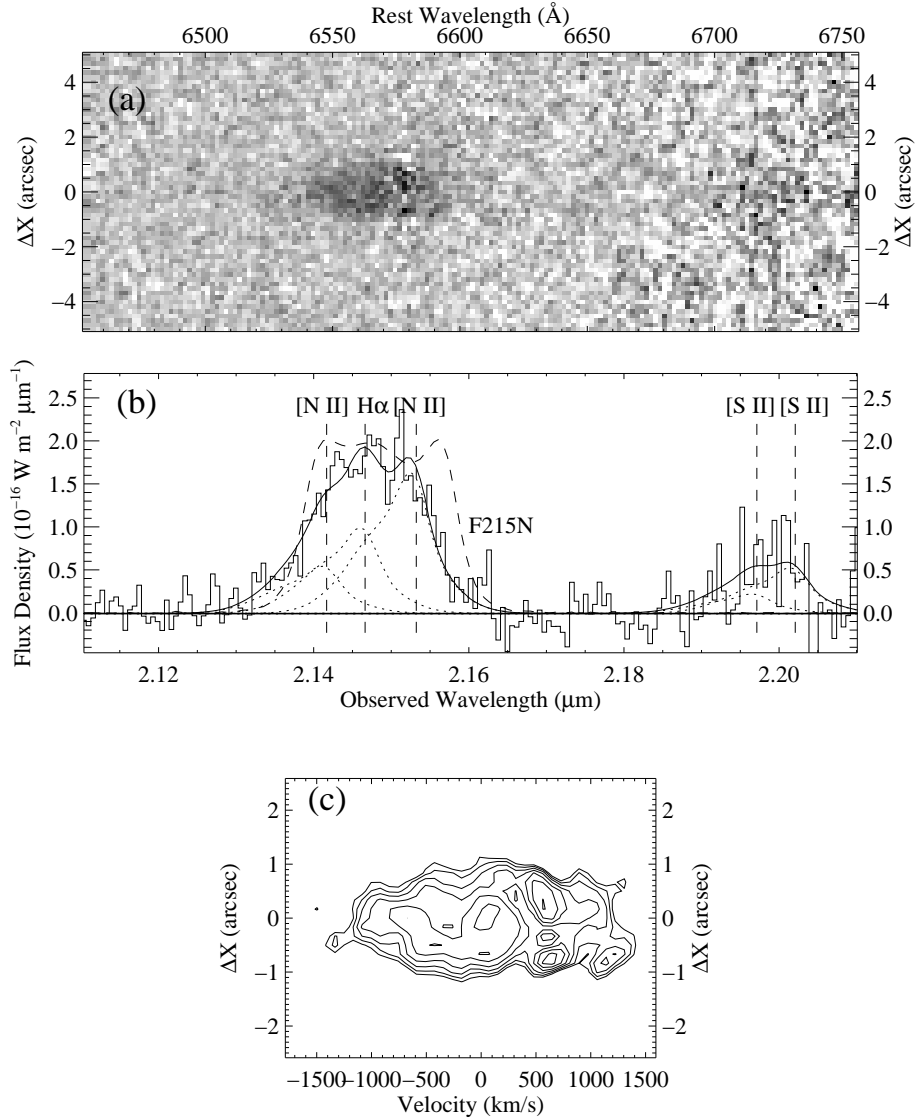,width=5.5in}

  \vspace*{-2cm}

  \figcaption[fig7.ps]{The \hanii\ spectrum of 4C~40.36 taken with the
  Palomar long-slit near-IR spectrograph: (a) The two dimensional
  image of the spectrum; (b) The summed one-dimensional spectrum.  The
  solid line shows our model fit with the five separate emission lines
  (\nii\ 6548/6583, \ha, \sii\ 6716/6731) shown by the dotted lines.
  Each emission line was modeled with two velocity components (see the
  text).  The dashed curve is the transmission of the HST/NIC2 F215N
  filter; (c) The intensity contours of the \hanii\ lines as a
  function of position and line-of-sight velocity.  The spectral image
  (a) was smoothed with a 2-pix FWHM Gaussian.  In the panels (a) and
  (c), the $-\Delta$X direction corresponds to a position angle of
  84$\degr$ (E of N).  \label{sp_ha}}

\end{figure}

\begin{figure}

  \vspace*{-1cm}

  \hspace*{1cm}\psfig{file=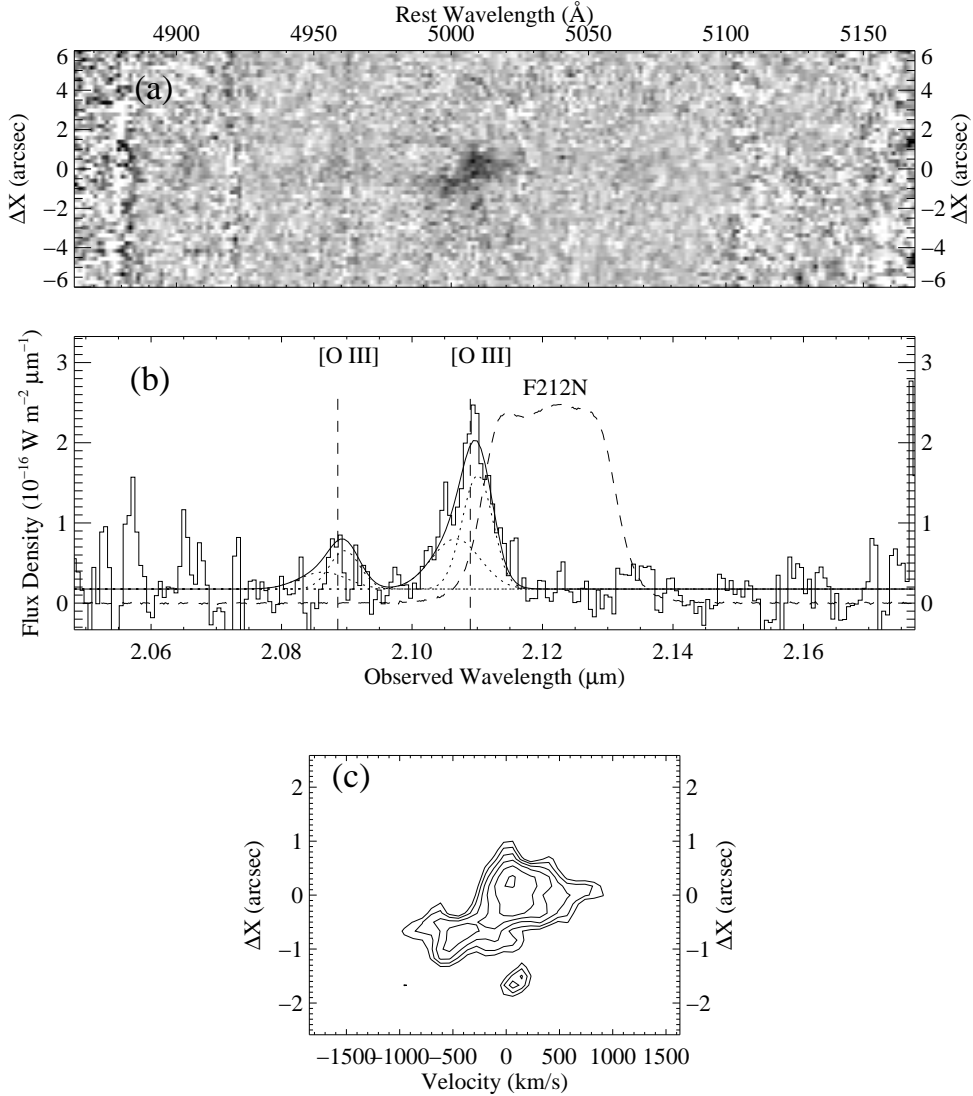,width=5.5in}

  \vspace*{-2cm}

  \figcaption[fig8.ps]{The \oiii\ spectrum of 4C~39.37 taken with
  the Palomar long-slit near-IR spectrograph: (a) The two dimensional
  image of the spectrum; (b) The summed one-dimensional spectrum.  The
  solid line shows our model fit with two velocity components shown
  with the dotted lines (see the text). The dashed line shows the
  transmission of the NICMOS F212N filter; (c) The intensity contours
  of the \oiii\ 5007\AA\ line as a function of position and
  line-of-sight velocity.  The spectral image (a) was smoothed with a
  2-pix FWHM Gaussian.  The two spatially distinct components
  correspond to the two velocity components in (b).  In the panels (a)
  and (c), the $-\Delta$X direction corresponds to a position angle of
  140$\degr$ (E of N).  \label{sp_o3b}}

\end{figure}

\begin{figure}

  \psfig{file=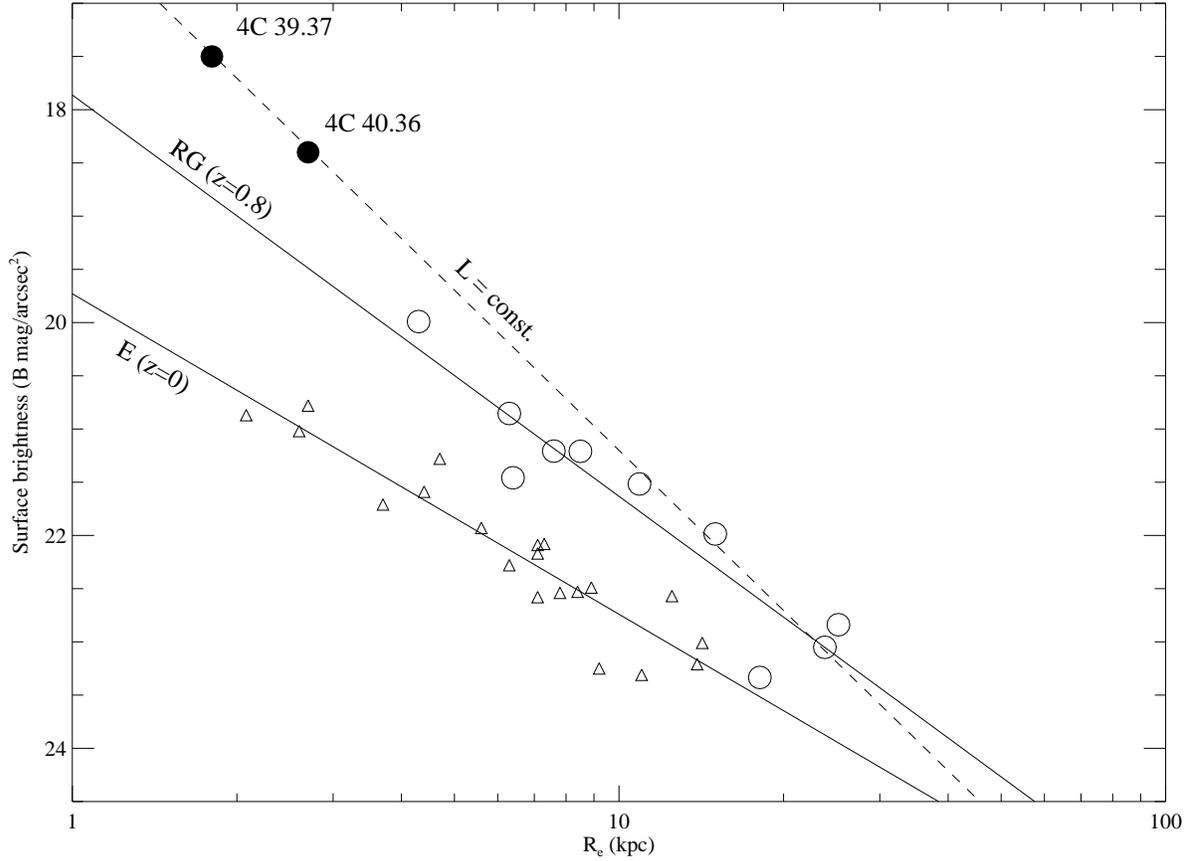,width=6.5in,angle=90}

  \figcaption[fig9.ps]{The effective radius and surface brightness of
  4C~40.36 and 4C~39.37 (solid circles) are compared with those of $z
  \sim 0.8$ radio galaxies (open circles, \citet{McLure00}) and $z
  \sim 0$ early-type galaxies (triangles, \citet{Kormendy77}).  The
  $I_{c}$-band surface brightnesses of \citet{McLure00} were
  transformed into the $B$ band by correcting for the $(1+z)^{3}$
  cosmological dimming (in $f_{\nu}$) only.  This is justified since
  the $I_{c}$ band samples the restframe $B$-band light of the $z \sim
  0.8$ radio galaxies.  The two solid lines indicate the Kormendy
  relation defined by the $z \sim 0.8$ radio galaxies and by the $z=0$
  early-type galaxies while the dashed line is the relation for
  constant luminosity galaxies ($L \propto r_{e}^{2} I_{e}$) scaled
  such that it crosses the two HzRG points.  \label{kormendy}}

\end{figure}

\begin{figure}

  \hspace*{1in}\psfig{file=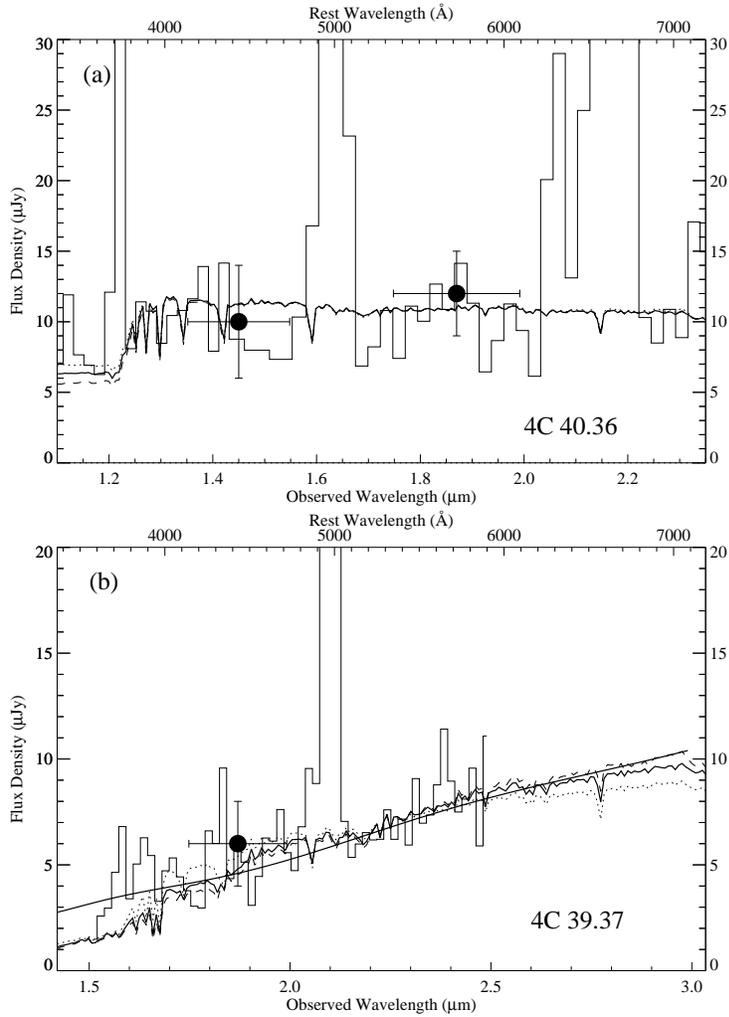,width=4in}

  \figcaption[fig10.ps]{Model fits to the NIRC spectra of 4C~40.36 (a)
  and 4C~39.37 (b), further rebinned by 4 and 3 pixels, respectively.
  The solid circles represent flux densities in the HST filters with
  the horizontal error bars showing the filter passbands---(a) The
  dotted, thick-solid, and dashed lines show the GISSEL96
  instantaneous burst models with an age of (45, 64, 90) Myrs and a
  mass of (3.0, 3.8, 4.7) $\times 10^{10}$ M$_{\odot}$, respectively;
  (b) The dotted, thick-solid, and dashed lines show the GISSEL
  instantaneous burst models with an age of (0.7, 1.0, 1.4) Gyrs and a
  mass of (1.6, 2.2, 3.3) $\times 10^{11}$ M$_{\odot}$, respectively.
  The thick solid line shows a flat ($f_{\nu} = 30 \mu$Jy) spectrum
  reddened by the extinction law of \citet{Cardelli89} with
  $E(B-V)=0.5$ mag. \label{lowspec}}

\end{figure}


\begin{thebibliography}{}

\bibitem[Archibald et al.(2001)]{Archibald01} Archibald, E.\ N., 
Dunlop, J.\ S., Hughes, D.\ H., Rawlings, S., Eales, S.\ A., \& Ivison, R.\ 
J.\ 2001, \mnras, 323, 417 

\bibitem[Armus et al.(1998)]{Armus98} Armus, L., Soifer, B.\ 
T., Murphy, T.\ W., Neugebauer, G., Evans, A.\ S., \& Matthews, K.\ 1998, 
\apj, 495, 276 

\bibitem[Bruzual \& Charlot(1993)]{Bruzual93} Bruzual, G. A., \&
Charlot, S. 1993, \apj, 405, 538

\bibitem[Cardelli, Clayton, \& Mathis(1989)]{Cardelli89} Cardelli, 
J.\ A., Clayton, G.\ C., \& Mathis, J.\ S.\ 1989, \apj, 345, 245 

\bibitem[Carilli et al.(1997)]{Carilli97} Carilli, C. L.,
R\"{o}ttgering, H. J. A., van Ojik, R., Miley, G. K., \& van Breugel,
W. J. M. 1997, \apjs, 109, 1

\bibitem[Carson et al.(2001)]{Carson01} Carson, J.~E.~et
al.\ 2001, \apj, 563, 63

\bibitem[Chambers, Miley, \& van Breugel(1988)]{Chambers88} Chambers, 
K. C., Miley, G. K. \& van Breugel, W. J. M. 1988, \apjl, 327, L47 

\bibitem[Chambers et al.(1996)]{Chambers96} Chambers, K. C., Miley, 
G. K., van Breugel, W. J. M., Bremer, M. A. R., Huang, J. -S. \& Trentham, 
N. A. 1996, \apjs, 106, 247 

\bibitem[De Breuck, et al.(2002)]{Breuck02} De Breuck, C., van
Breugel, W., Stanford, A., Roettgering, H., J. A., Miley, G., \&
Stern, D. 2002, \aj, 123, 637

\bibitem[Eales \& Rawlings(1993a)]{Eales93} Eales, S. A. \& 
Rawlings, S.  1993a, \apj, 411, 67 

\bibitem[Eales \& Rawlings(1996)]{Eales96} Eales, S.\ A.\ \& Rawlings,
S.\ 1996, \apj, 460, 68

\bibitem[Eales et al.(1993b)]{Eales93b} Eales, S. A., Rawlings, S.,
Dickinson, M., Spinrad, H., Hill, G. J., \& Lacy, M. 1993b, \apj, 409, 578

\bibitem[Egami et al.(1999)]{Egami99} Egami, E., Arumus, L., 
Neugebauer, G., Soifer, B.\ T., Evans, A.\ S., \& Murphy, T.\ W.\ 1999, ASP 
Conf.\ Ser.\ 193: The Hy-Redshift Universe: Galaxy Formation and Evolution 
at High Redshift, 86 

\bibitem[Evans(1998)]{Evans98} Evans, A. S. 1998, \apj, 498, 553 

\bibitem[Fruchter \& Hook(2002)]{Fruchter02} Fruchter, A. S., \& Hook,
R. N. 2002, \pasp, 114, 144

\bibitem[Hawarden et al.(2001)]{Hawarden01} Hawarden, T. G., Leggett,
S. K., Letawsky, M. B., Ballantyne, D. R., \& Casali, M. M. 2001,
\mnras, 325, 563

\bibitem[Hunt et al.(1998)]{Hunt98} Hunt, L. K., Mannucci, F., Testi,
L., Migliorini, S., Stanga, R. M., Baffa, C., Lisi, F., \& Vanzi,
L. 1998, \aj, 115, 2594

\bibitem[Iwamuro et al.(1996)]{Iwamuro96} 
Iwamuro, F., Oya, S., Tsukamoto, H. \& Maihara, T. 1996, \apjl, 466, L67 

\bibitem[Jarvis et al.(2001)]{Jarvis01} Jarvis, M.~J.,
Rawlings, S., Eales, S., Blundell, K.~M., Bunker, A.~J., Croft, S.,
McLure, R.~J., \& Willott, C.~J.\ 2001, \mnras, 326, 1585

\bibitem[Kormendy(1977)]{Kormendy77} Kormendy, J. 1977, \apj, 218, 333

\bibitem[Larkin et al.(1996)]{Larkin96} Larkin, J. E., Knop, R. A.,
Lin, S., Matthews, K., \& Soifer, B. T. 1996, 108, 211

\bibitem[Leitherer et al.(1999)]{Leitherer99} Leitherer, C. , et 
al. 1999, \apjs, 123, 3 

\bibitem[Lejeune, Cuisinier and Buser(1997)]{Lejeune97} Lejeune, 
T., Cuisinier, F. and Buser, R. 1997, \aaps, 125, 229 

\bibitem[Lilly \& Longair(1984)]{Lilly84} Lilly, S. J. and 
Longair, M. S. 1984, \mnras, 211, 833 

\bibitem[McLure \& Dunlop(2000)]{McLure00} McLure, R.\ J.\ \& 
Dunlop, J.\ S.\ 2000, \mnras, 317, 249 

\bibitem[Motohara et al.(2001)]{Motohara01} Motohara, K. et al. 2001,
\pasj, 53, 459

\bibitem[Pentericci et al.(2001)]{Pentericci01} Pentericci, L., 
McCarthy, P.\ J., R{\" o}ttgering, H.\ J.\ A., Miley, G.\ K., van Breugel, 
W.\ J.\ M., \& Fosbury, R.\ 2001, \apjs, 135, 63 

\bibitem[Rawlings, Eales, \& Warren(1990)]{Rawlings90} Rawlings, 
S., Eales, S., \& Warren, S.\ 1990, \mnras, 243, 14

\bibitem[Rawlings et al.(1989)]{Rawlings89} Rawlings, S., Saunders,
R., Eales, S. A., \& Mackay, C. D. 1989, MNRAS, 240, 701

\bibitem[Simpson et al.(1999)]{Simpson99} Simpson, C. et al. 1999, \apj,
525, 659

\bibitem[Thompson et al.(1998)]{Thompson98} Thompson, R.~I.,
Rieke, M., Schneider, G., Hines, D.~C., \& Corbin, M.~R.\ 1998, \apjl,
492, L95

\bibitem[van Breugel et al.(1998)]{Breugel98} van Breugel, W. J. M.,
Stanford, S. A., Spinrad, H., Stern, D., \& Graham, J. R. 1998, \apj,
502, 614

\bibitem[Veilleux \& Osterbrock(1987)]{Veilleux87} Veilleux, S.\ 
\& Osterbrock, D.\ E.\ 1987, \apjs, 63, 295 

\bibitem[Vernet et al.(2001)]{Vernet01} Vernet, J., Fosbury, R.\ 
A.\ E., Villar-Mart{\'i}n, M., Cohen, M.\ H., Cimatti, A., di Serego 
Alighieri, S., \& Goodrich, R.\ W.\ 2001, \aap, 370, 407 

\bibitem[Willott et al.(1999)]{Willott99} Willott, C. J., Rawlings,
S., Blundell, K. M., \& Lacy, M. 1999, MNRAS, 309, 1017

\end{thebibliography}
\end{document}